\documentclass[conference]{IEEEtran}
\IEEEoverridecommandlockouts
\usepackage{cite}
\usepackage{amsmath,amssymb,amsfonts}
\usepackage{algorithmic}
\usepackage{graphicx}
\usepackage{textcomp}
\usepackage{xcolor}
\usepackage[colorlinks,
            linkcolor=blue,
            anchorcolor=blue,
            urlcolor=blue,
            citecolor=blue]{hyperref}
\usepackage{fancyhdr}
\def\BibTeX{{\rm B\kern-.05em{\sc i\kern-.025em b}\kern-.08em
    T\kern-.1667em\lower.7ex\hbox{E}\kern-.125emX}}

\makeatletter
\newcommand{\linebreakand}{%
  \end{@IEEEauthorhalign}
  \hfill\mbox{}\par
  \mbox{}\hfill\begin{@IEEEauthorhalign}
}
\makeatother

\fancypagestyle{whitepaper}{
    \fancyhf{}
    \fancyhead[C]{White Paper v1.1.0.1}
    \fancyfoot[C]{\thepage}
    
}

\begin{document}

\title{ DASICS: Enhancing Memory Protection with Dynamic Compartmentalization }


\author{
\IEEEauthorblockN{Yue Jin}
\and
\IEEEauthorblockN{Yibin Xu}
\and
\IEEEauthorblockN{Chengyuan Yang}
\and
\IEEEauthorblockN{Han Wang  }
\and
\IEEEauthorblockN{Tianyi Huang  }
\and
\IEEEauthorblockN{Tianyue Lu}
\and
\IEEEauthorblockN{Mingyu Chen}
\linebreakand
\IEEEauthorblockA{\textit{Institute of Computing Technology, Chinese Academy of Sciences} \\
Beijing, China }
}

\maketitle
\thispagestyle{whitepaper}


\begin{abstract}
In the existing software development ecosystem, security issues introduced by third-party code cannot be overlooked. Among these security concerns, memory access vulnerabilities stand out prominently, leading to risks such as the theft or tampering of sensitive data. To address this issue, software-based defense mechanisms have been established at the programming language, compiler, and operating system levels. However, as a trade-off, these mechanisms significantly reduce software execution efficiency. Hardware-software co-design approaches have sought to either construct entirely isolated trusted execution environments or attempt to partition security domains within the same address space. While such approaches enhance efficiency compared to pure software methods, they also encounter challenges related to granularity of protection, performance overhead, and portability.

In response to these challenges, we present the DASICS (Dynamic in-Address-Space Isolation by Code Segments) secure processor design, which offers dynamic and flexible security protection across multiple privilege levels, addressing data flow protection, control flow protection, and secure system calls. We have implemented hardware FPGA prototypes and software QEMU simulator prototypes based on DASICS, along with necessary modifications to system software for adaptability. We illustrate the protective mechanisms and effectiveness of DASICS with two practical examples and provide potential real-world use cases where DASICS could be applied.
\end{abstract}


\section{Introduction}
The open-source, shared, and collaborative software development model has significantly contributed to advancing fields such as the Internet and artificial intelligence. However, this model, characterized by multiple developers working together on a single software project, frequent reliance on third-party software components, and managing large code volumes, has increased software development complexity. This complexity has resulted in a higher probability of introducing security vulnerabilities during development. For example, software developers often need to incorporate third-party libraries, but the lack of security assurance of these libraries increases the risk of information leakage and tampering. Attackers can exploit the vulnerabilities in unreliable third-party libraries. Moreover, when frequently used third-party libraries contain such dangerous vulnerabilities, A vast number of software applications developed using these libraries are affected.

The primary vulnerability in software security is memory access vulnerability. To address these vulnerabilities, both academia and industry have proposed various software and hardware methods for memory protection. On one hand, these protection methods involve thorough inspection and restriction of data flow in non-trusted software code using data flow integrity technology (DFI). This entails boundary checks on data and compliance checks on data sources, aiming to prevent unauthorized memory operations. Representative works in this field encompass Intel's MPX\cite{MPX} and MPK technologies\cite{MPK}, ARM's MTE technology\cite{ARM-MTE}, as well as the CHERI\cite{watson2015cheri} security architecture developed by Cambridge. Additionally, control flow integrity technology (CFI) is employed to thwart malicious control flow hijacking, with implementations like Intel's CET technology\cite{CET}, ARM's BTI\cite{ARM-BTI} technology, and Pointer Authentication\cite{ARM-PA}. 

Nevertheless, existing memory protection methods face challenges of different magnitudes, including issues of coarse-grained object isolation, vulnerability of security metadata to attacks, hardware implementation/performance costs, and the requirement for substantial modifications and recompilation of existing third-party code.

We propose DASICS, a security architecture design to address these problems of coarse granularity of isolated objects, low metadata security, and excessive performance overhead of existing security protection techniques, and focus on dynamic compartmentalizing, memory protection within the same level of address space, and cross-level call checking, which have received less attention in previous work. We implement DASICS based on code segmentation to do dynamic compartmentalization of permission domains, provide hardware-assisted efficient software memory protection, guarantee the safe invocation and operation of third-party code, and provide solid security assurance and support for open source and open software development.

\section{Background}

Today's software ecosystem heavily relies on foundational software components, including operating systems and network protocol stacks, as well as high-performance third-party libraries such as SQLite, ASL, Crypto++, and Qt, which are predominantly implemented in C/C++ languages. However, it is critical to acknowledge that many of these third-party libraries may harbor memory access vulnerabilities.


Regrettably, software developers who leverage these third-party libraries often overlook non-security aspects of the library code itself, such as neglecting to perform essential boundary checks. Consequently, the software applications built upon these libraries inherit security vulnerabilities that can be maliciously exploited and eventually leading to severe security implications, including but not limited to data leakage and tampering.

Memory access vulnerabilities constitute a significant and prevalent threat to software security today\cite{szekeres2013sok}. These vulnerabilities occur when a program erroneously accesses or alters memory outside its authorized bounds, either due to programming errors or malicious attacks. Such incidents can disrupt normal program execution, potentially leading to system crashes or exploitation by attackers. Common memory access vulnerabilities encompass buffer overflows\cite{cowan2000buffer}, uninitialized memory usage\cite{stepanov2015memorysanitizer}, null pointer references\cite{meyer2017ending}, heap overflows\cite{jia2017towards}, etc. These vulnerabilities can result in program crashes, data loss, system instability, information leakage, and other security concerns, thereby posing substantial risks to overall system integrity.

\begin{figure}
\centering
\includegraphics[width=80mm]{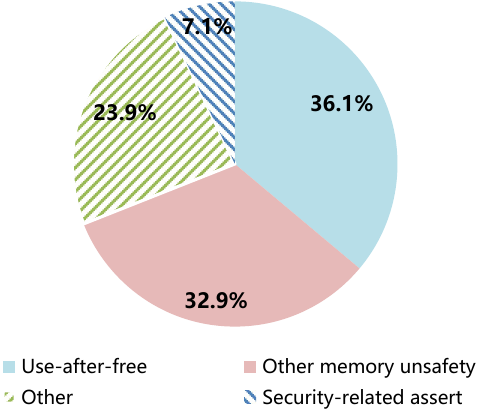}
\caption{Statics of Chrome Security Vulnerabilities}
\label{fig:chrome}
\end{figure}

Figure \ref{fig:chrome} illustrates the origins of challenging security vulnerabilities in the Chrome codebase based on Google's statistics\cite{Chrome}. Notably, 36.1\% of these vulnerabilities are attributable to Use-after-free attacks, primarily stemming from improper memory management and reuse after memory release. Additionally, 32.9\% originate from various other security issues, including overflow attacks. In terms of statistical analysis, it is noteworthy that approximately 69\% of the identified vulnerabilities fall into the category of memory access vulnerabilities.

Memory access vulnerabilities in third-party library code typically stem from two main sources: the absence of boundary checks on data accessible to non-trusted code and the lack of control flow checking within the program. Consider the Heartbleed vulnerability (CVE-2014-0160) that emerged in OpenSSL prior to version 1.0.1. In this case, the library failed to validate the input length (lacking bounds checking) during the implementation of the heartbeat detection mechanism in the Transport Layer Security (TLS) protocol. This oversight resulted in a vulnerability exploitable by attackers to conduct Overread Buffer attacks on both clients and servers. The consequences were severe, including critical information leakage and widespread impacts\cite{durumeric2014matter,ghafoor2014analysis,wang2015risk}.

In more critical memory vulnerabilities, attackers can exploit the flaw to alter the instruction execution address, thereby redirecting the program's control flow to malicious code under their control. This opens the door to a wider range of flexible and dangerous attacks, including those with Turing-complete capabilities\cite{roemer2012return,prandini2012return,hu2016data}.   

Third-party code libraries often share the same process address space as the code developed by software developers, rendering vulnerabilities in third-party code a potential security threat to the entire program. In programming languages like C/C++, lacking built-in memory security checks, each line of code theoretically enjoys unrestricted access to any memory location within the process address space. When an attacker exploits a vulnerability in third-party code, they can gain access to any data and code within the process address space, enabling data theft, tampering, program hijacking, and potential further exploitation of system resources. In particular, a vulnerability in a third-party driver within the OS kernel can provide unfettered access to the system's entire memory.

\section{Rethinking of Existing Methodology}
Existing research in the field of memory access vulnerability protection can be broadly categorized into two primary approaches. First, there is an emphasis on efficient software partitioning and subsequent boundary checking from a data flow perspective. This approach involves partitioning the software in a way that constrains certain sections of code to access only their designated data, preventing unauthorized access to data beyond their bounds. For instance, a function is prevented from modifying local variables of higher-level functions due to stack overflow protection measures.

The second approach focuses on control flow restrictions. It involves monitoring control flow and triggering exceptions if a function attempts to return to an address different from its original call address, thus preventing unauthorized jumps. Different security methodologies offer varying degrees of completeness, user-friendliness, and performance overhead. The selection of an appropriate system design must carefully balance these factors.

Defense mechanisms against memory security problems can be classified as pure software approaches and hardware-software collaborative approaches. The software-only approach can be protected at different levels, such as programming languages, compilers, and loaders. The hardware-software approach includes the construction of a trusted execution environment for program operation, as well as address space isolation and control flow protection within the program.

\subsection{Software Defenses}

A purely software-based defense strategy has the potential to provide comprehensive protection across all levels. From a programming language perspective, addressing memory access vulnerabilities can be effectively mitigated by adopting memory-safe languages, such as Rust\cite{matsakis2014rust}, to reimplement untrusted code. Nevertheless, this approach encounters substantial practical challenges. The formidable engineering undertaking involved in translating a diverse array of third-party libraries, replete with intricate codebases, from languages like C/C++ to a secure language like Rust, is considerably daunting.

From the compiler's perspective, safeguarding against runtime out-of-bounds behavior entails the insertion of real-time verification code into the source code. One illustrative technique is the Stack Protector\cite{dang2015performance}, already integrated into GCC and Clang compilers\cite{StackProtector}. This technique primarily fortifies against stack-based attacks by establishing a stack protection variable during function invocation and subsequently verifying its integrity upon function return. Another notable approach is the AddressSanitizer (ASan) \cite{serebryany2012addresssanitizer} compiler plug-in, supported by the Microsoft C/C++ compiler. ASan achieves its objectives by injecting detection code into the target program during compilation, thereby enabling the runtime detection of memory-related errors.

Address Space Layout Randomization (ASLR)\cite{marco2019address,lu2015aslr} represents a crucial security mechanism aimed at thwarting code-jumping attacks, particularly those that leverage memory vulnerabilities. Its fundamental premise revolves around the deliberate introduction of randomness into the organization of linear memory regions, including the heap, stack, and mappings of shared libraries. This strategic randomization hinders malicious programs from pinpointing the precise location of target code segments, thus significantly augmenting the complexity of predicting destination addresses for potential attackers.

The safeguarding of software control flow hinges on the Control Flow Integrity (CFI) technique\cite{abadi2009control,burow2017control}. Initially, CFI constructs a representation of a program's Control Flow Graph (CFG) through static analysis of either the program's source code or binary code. Subsequently, it monitors the program's runtime execution to ascertain whether it falls victim to attacks. This monitoring process involves verifying that the program's control flow consistently adheres to the CFG.

The Control Flow Graph (CFG) is graphically depicted as a collection of basic blocks interconnected by directed edges. Each basic block corresponds to a continuous segment of code devoid of control flow jump instructions and is not the target of any other control flow jump instructions. Conversely, directed edges signify transitions in the program's control flow, signifying the movement of execution from one basic block to another.

Although the pure software approach has demonstrated its capability to offer certain security defenses, it is essential to acknowledge that the incorporation of techniques like code instrumentation (e.g., methods akin to ASan) and the insertion of validation code prior to and after memory access incurs a notable performance overhead on the original software operations. Additionally, security mechanisms that necessitate the safeguarding of security metadata, such as CFG check methods, must ensure the integrity and authenticity of these security metadata. Ensuring that these metadata used for security assessments remain untampered and unaltered by potential attackers becomes a crucial concern in such cases.

\subsection{Trusted Execution Environment}
The Trusted Execution Environment (TEE) constitutes a security architecture rooted in the collaborative synergy of hardware and software components. Its fundamental purpose revolves around the establishment of a secure computational realm, shielded from external threats, achieved through mechanisms such as time-division multiplexing of CPUs or partitioning specific memory addresses to carve out a secure enclave. This fortified environment effectively safeguards against unauthorized access, reading, or tampering of internal data and control flow by malicious programs from external sources, courtesy of robust hardware-backed isolation, encryption, and integrity verification measures.

The utility of TEE extends across various domains, including mobile devices and cloud computing, where it plays a pivotal role in fortifying security. Prominent technologies embodying TEE principles include Intel's SGX\cite{SGX} security technology, AMD's SEV-SNP\cite{SEV-SNP} secure virtualization technology, CCA confidential computing technology\cite{ARM-CCA}, and ARM architecture's TrustZone\cite{ARM-Trustzone} processor security isolation technology. Additionally, architectures like RISC-V feature security solutions such as Penglai\cite{feng2021scalable} and Keystone\cite{lee2020keystone}, which adhere to the TEE paradigm.

Specific TEE technologies are inherently tied to particular architectures or platforms, which can lead to increased costs associated with hardware replacement and software migration efforts. Furthermore, to maintain the security of the computational environment, TEE implementations often necessitate a substantial allocation of hardware resources, including measures like securing memory data integrity using structures like Merkle trees\cite{feng2021scalable}.

Typically, there exists a notable performance overhead when dealing with interactions between data inside and outside the TEE. Additionally, the overhead associated with creating and switching between secure environments cannot be disregarded. Consequently, TEE finds its optimal utility in scenarios where relatively independent, small-scale security core code is run, while it may be less suitable for use cases requiring frequent interactions.

In theory, it is conceivable to run extensive applications entirely within a TEE; however, TEE functionality may not adequately address the complexities of isolating multiple modules within a large program, especially when dealing with third-party libraries.

\subsection{Compartmentalization}
\begin{figure}
\centering
\includegraphics[width=88mm]{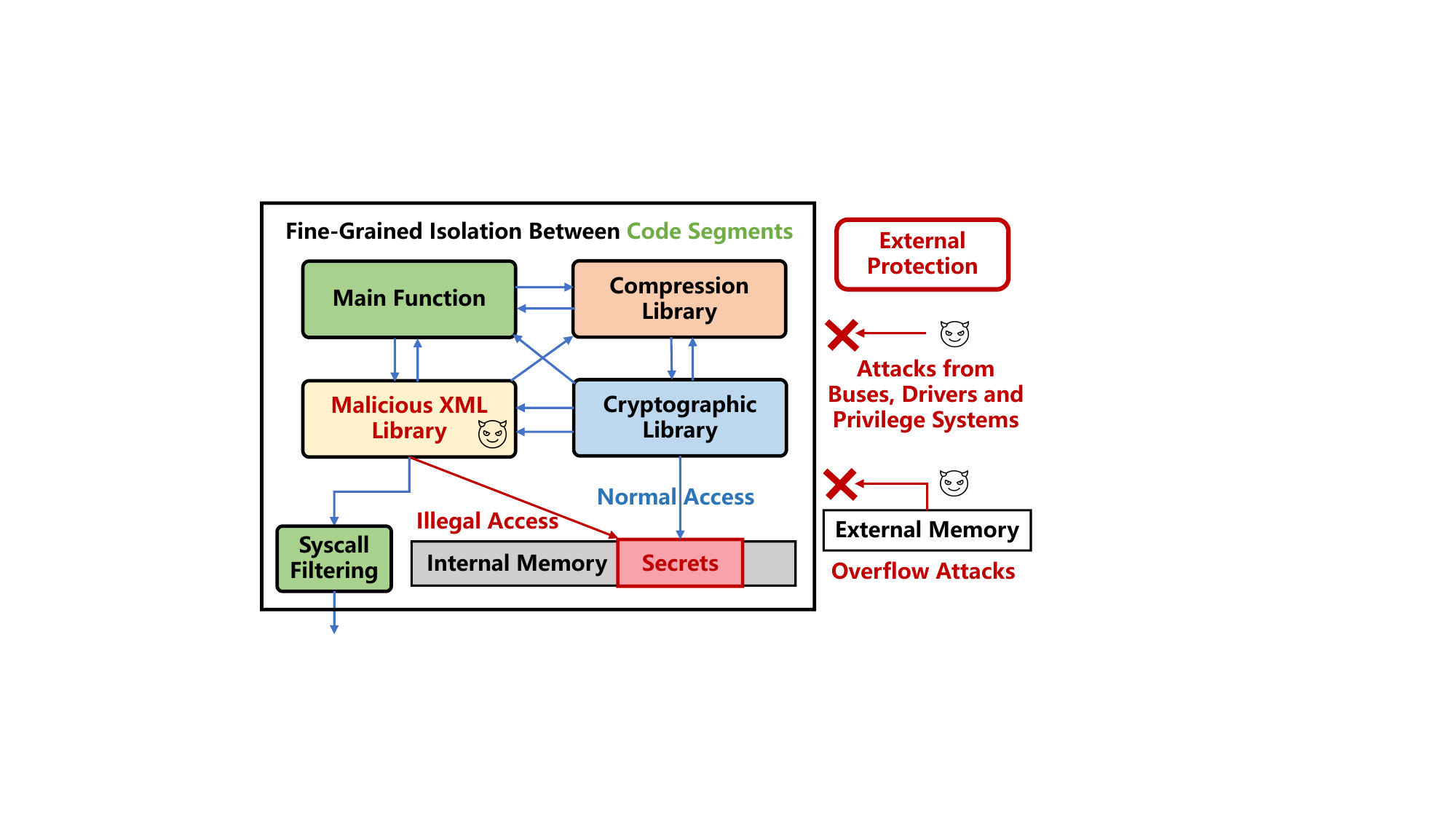}
\caption{Threats from Non-Trusted Libraries within Software}
\label{fig:internal}
\end{figure}

Beyond ensuring isolation and protection against external threats through mechanisms like TEE, addressing vulnerabilities posed by threats originating within the software, such as non-trusted third-party code bases, is imperative. In scenarios where there is no inherent isolation among code segments within the software and all code possesses identical privileges with unrestricted access to the address space—enabling access to any data and the ability to jump to any code within that space—potential risks arise. Non-trusted code could exploit this scenario to pilfer sensitive data, like user passwords, handled by the software. Moreover, it could potentially execute malicious attack code through unauthorized jumps, leading to more severe and damaging assaults. Consequently, it becomes essential to delineate and segregate access and jump domains based on distinct code segments within the software. (Figure \ref{fig:internal})

Internal protection mechanisms within software must encompass two key aspects: the segregation of permissions for individual compartments within the software and the imposition of constraints on memory access and jump destinations for non-trusted third-party libraries. These measures are essential to counteract unauthorized out-of-bounds access, tampering with sensitive data, and the prevention of control flow hijacking that could lead to the execution of malicious code. Consequently, protection techniques within the software can be categorically organized based on their roles in ensuring both data access isolation and control flow safeguarding.

\subsubsection{\textbf{Data flow integrity}}

The concept of Data Flow Integrity (DFI) primarily revolves around the partitioning and isolation of data that various sections of software are permitted to access internally. Simultaneously, during memory accesses, DFI entails scrutinizing these accesses based on the established isolation rules to prevent out-of-bounds access and illicit operations. The overarching objective is to thwart unauthorized data theft and tampering, particularly when dealing with sensitive and critical information. Depending on the specific partitioning approach, DFI methodologies can be further categorized into those that extend permissions to pointers and those that extend permissions to memory data.

Representative technologies that focus on extending permissions to pointers include MPX and CHERI. The central idea behind these approaches involves defining the permissible range of memory addresses for each pointer and specifying the operations allowed on the data pointed to by the pointer. During memory accesses involving pointers, these defined rules are employed to conduct checks and ensure compliance.

MPX (Memory Protection Extensions)\cite{MPX} is a hardware extension technology developed by Intel to mitigate memory access vulnerabilities. MPX introduces four 128-bit bound registers, four boundary address setting and comparison instructions, four boundary register movement instructions, two configuration registers, and one status register. During program execution, MPX enhances code security by examining pointer references to determine whether they exceed predefined boundaries. Due to the limited number of bound registers, additional pointer-bound metadata must be stored in a separate table in memory. However, the adoption of this technology presents significant challenges. It necessitates substantial software modifications, including recompilation of third-party libraries, incurring substantial software development costs. Furthermore, the need for code instrumentation to inspect each memory access instruction leads to severe performance degradation. These issues collectively prompted Intel to remove MPX technology from the x86 architecture in 2019.

CHERI (Capability Hardware Enhanced RISC Instructions) \cite{watson2015cheri,woodruff_cheri_2014,watson_fast_2016,wesley_filardo_cornucopia_2020,chisnall_cheri_2017 }represents another memory protection technology that operates through a synergy of hardware and software components. CHERI technology extends the semantics of pointers, transforming conventional address-length pointers into 128-bit Extended Capability pointers (Capabilities) that encompass information about the pointer's valid range, permissions, and validity status. Specialized instructions are introduced to manipulate the information within these extended pointers.CHERI holds the potential to provide memory security protection at the pointer level, theoretically capable of defending against most overflow attacks. However, the adoption of CHERI comes with substantial memory access overhead due to the extended pointers. Additionally, it mandates the insertion of instructions to modify these extended pointers, necessitating extensive modifications to compilers and the recompilation of third-party library code.

Representative technologies for extending permissions to memory data include MTE\cite{ARM-MTE} and PKU, among others. These technologies primarily implement data isolation by introducing memory data region tagging (tagging). Code, in turn, gains explicit access permissions to specific memory regions by virtue of holding these tags.

The Memory Tagging Extension (MTE) technology\cite{ARM-MTE} is an introduced memory protection mechanism in the ARMv8.5-A architecture. It divides memory into multiple 64-byte sections, assigning a unique tag to each section. These tags are utilized in conjunction with pointers to verify whether the pointers correctly reference the intended memory locations. MTE primarily enforces memory protection through the use of these hardware-based tags without necessitating any code modifications. Furthermore, MTE offers software-based tagging capabilities, allowing applications to customize tags through various tagging schemes, including address-based and random-based tagging. It also provides multiple tag width options to strike a balance between the number of tags and the additional overhead incurred by tagging.

PKU (Protection Key Unit) technology is a lightweight memory isolation mechanism exemplified by Intel's Memory Protection Key (MPK) technology\cite{MPK, schrammel2020donky,gu_epk_2022,park_libmpk_2019,vahldiek-oberwagner_erim_2019}. It partitions memory regions at a page granularity by assigning a "color" to each page, utilizing the extra 4 bits available in page table entries as key values. MPK introduces a dedicated register, PKRU (Protection Key Rights Register), to maintain the current program's key ("color"), allowing it to detect illegal memory accesses when the accessed page does not permit the current key's access. MPK also incorporates special instructions for manipulating PKRU, enabling memory isolation with minimal hardware overhead. However, MPK lacks an integrated security mechanism to safeguard PKRU from tampering. Consequently, it necessitates inspecting binary library functions to ensure they do not modify PKRU instructions. Furthermore, MPK supports a maximum of 16 keys, limiting the concurrent existence of only 16 keys. This limitation proves insufficient for complex programs with substantial isolation requirements.

\subsubsection{\textbf{Control flow integrity}}
Control Flow Integrity (CFI) technology\cite{abadi2009control,burow2017control} originally required the construction of a comprehensive representation of program control flow, such as a CFG. However, in more practical methods, a common approach is to protect critical control flow data elements, such as function pointers and return addresses stored on the stack, in an efficient and straightforward manner. Additionally, CFI techniques aim to restrict target addresses for indirect jumps, often limited to specific locations like function entry points. Key technologies in this domain include CET (Control-flow Enforcement Technology)\cite{CET}, BTI (Branch Target Identification), and PA (Pointer Authentication), among others.

CET (Control-flow Enforcement Technology)\cite{CET} is a novel security technology developed by Intel. It primarily focuses on safeguarding control flow at the hardware level, effectively preventing hijacking and tampering attempts targeting control flow. CET encompasses two specific mechanisms:
\begin{itemize}
    \item \textbf{\textit{Indirect Branch Tracking (IBT)}}: This mechanism primarily addresses attacks like JOP (Jump-Oriented-Programming)\cite{bletsch2011jump} and ROP (Return-Oriented-Programming)\cite{prandini2012return}, which exploit indirect jumps. IBT works by setting landing-point instructions for indirect jump instructions, constraining legitimate indirect jump target addresses within the program, such as function entry points, to ensure the legitimacy of jump destinations.
    \item \textbf{\textit{Shadow Stack (SS)}}: SS operates by maintaining an additional stack to track return addresses during function calls. It involves pushing return addresses onto the stack during function calls and subsequently popping and comparing return addresses with the target of the jump when returning from a function. This process effectively thwarts attackers from tampering with return addresses, enhancing security.
\end{itemize}


BTI (Branch Target Identification)\cite{ARM-BTI} is a security feature introduced in ARMv8.5, specifically designed to counteract attacks like JOP (Jump-Oriented-Programming)\cite{bletsch2011jump}. Similar to IBT (Indirect Branch Tracking), BTI operates during translation time by placing BTI instructions at specific locations for indirect jump instructions. This approach serves to define the legitimate targets for indirect jumps, making it more challenging for JOP attacks to string together program execution through indirect jumps.

Pointer Authentication (PA) technology\cite{ARM-PA} is an integrity verification technique developed by ARM. PA employs cryptographic integrity measures to protect return addresses. In line with hardware resource conservation principles, PA leverages unused high-order bits within virtual address pointers. In collaboration with the compiler, the called function initially computes a checksum or authentication code for the return address and embeds it in the high-order bits of the pointer, known as the Pointer Authentication Code (PAC). This modified return address with the authentication code is then pushed onto the stack. Prior to a function return, verification instructions are inserted. If a discrepancy is detected between the verification result and the PAC, it indicates malicious tampering with the return address, triggering an exception. This approach enhances the security of return address protection.

\subsubsection{\textbf{Metadata protection}}

Existing address space isolation methods share a common issue: some protective mechanisms rely on security metadata (such as MTE's tags and PKU's keys) for defense but lack protection for the security metadata itself. This vulnerability allows malicious programs to elevate their privileges or circumvent isolation constraints by altering the security metadata, thereby rendering the protective mechanisms ineffective. For example, in the case of the MPK protection mechanism, a program can employ instructions to freely manipulate the PKRU register, thereby rendering the coloring protection applied to itself ineffective.

In addressing the security metadata protection issue, numerous research endeavors have sought improvements from different perspectives, broadly falling into two directions:

\begin{itemize}
    \item \textbf{\textit{Binary Scanning}}: This method primarily involves inspecting and modifying the binary code of applications or non-trusted libraries to ensure that non-trusted code cannot contain instructions (often specially designed instructions) that alter security metadata. For instance, an academic solution like ERIM\cite{vahldiek-oberwagner_erim_2019} employs binary scanning to prevent third-party library code from including instructions like WRPKRU, which modify the PKRU register. However, it's worth noting that this approach may not adequately support dynamically generated code.
    \item \textbf{\textit{User-Level Privileged Function}}: This approach revolves around the concept of constructing a privileged function in user mode, dedicated to managing and manipulating security metadata. Any security metadata operations conducted outside of this function trigger exceptions. For instance, consider the academic solution Donky\cite{schrammel2020donky}: Donky utilizes user-mode interrupts to establish a security management function called the "Domain Monitor" in user mode. All modifications to PKRU can only occur within this function, and attempts to use the WRPKRU instruction outside of it result in exceptions. Additionally, Donky implements control-flow restrictions to prevent unauthorized transitions to the Domain Monitor, thus thwarting attempts to alter PKRU through control-flow hijacking. This approach resembles a form of privilege-level partitioning within user mode. However, even lightweight privilege-level partitioning may introduce performance overhead due to frequent privilege-level switches.
\end{itemize}

The protection of security metadata often necessitates a combined approach involving Data Flow Integrity (DFI) and Control Flow Integrity (CFI). Firstly, these metadata, being sensitive data themselves, require the security isolation provided by DFI methods to ensure that non-trusted code cannot access or tamper with them. Secondly, CFI methods are essential to guarantee that code with the capability to modify this sensitive data is not susceptible to hijacking or exploitation for malicious alterations.

\subsection{Common Issues in Designing Security Methods}
We have summarized the security mechanisms for mitigating memory access vulnerabilities mentioned above and believe that as these mechanisms transition into practical deployment, they must address the following real-world challenges:

\subsubsection {\textbf{The tradeoff between protection and overhead}}
Security solutions inevitably faces a trade-off between performance/hardware overhead and protection capability. On the one hand, certain mechanisms with strong protective capabilities may introduce a significant performance overhead due to frequent checks (e.g., MPX technology necessitates permission table queries for every pointer access, resulting in substantial performance costs). Security protections relying on privilege-level checks may require numerous privilege-level switches, also incurring noticeable performance overhead. On the other hand, some security mechanisms, particularly those involving hardware-software cooperation, rely on additional security metadata, which can lead to substantial hardware overhead. For example, CHERI extends pointers, significantly increasing the storage overhead of pointer variables. This can result in considerable storage and access overhead, especially for applications with a large number of pointers. Therefore, when designing security mechanisms, it is crucial to consider not only security aspects but also the cost required to achieve the desired level of security.

\subsubsection{\textbf{Portability}}
Some security mechanisms face challenges in terms of rewrite complexity and compilation difficulties when protecting existing third-party code libraries. This raises issues concerning the portability of these existing security mechanisms. Many academic efforts aim to achieve safety goals by introducing new instructions or extending pointers. However, these alterations change the program's runtime instructions and pointer semantics, resulting in incompatibility with legacy binary code libraries. Utilizing these security mechanisms for protection would require the recompilation of all non-secure third-party libraries. Besides the considerable workload involved, many third-party libraries are not open-source, making it challenging to apply these security mechanisms widely. In summary, when designing security mechanisms, portability concerns also need to be taken into account.

\subsubsection{\textbf{Security check for cross-privilege switching}}
DFI and CFI mechanisms are typically employed within the same privilege level to enhance fine-grained protection. However, once privilege-level switches occur, code at a higher privilege level often gains the privilege to modify code and data at lower privilege levels. Therefore, without imposing certain restrictions on privilege-level switches, the protective mechanisms of DFI and CFI could be compromised through privileged calls. Common privilege-level calls include system calls and virtual machine traps, among others. System calls serve as critical interfaces through which applications access functionality provided by the operating system. However, they can also be exploited by non-trusted code to bypass security mechanisms. For instance, under the MPK protection mechanism, code can use the \textit{madvise} system call (originally intended to provide additional memory access information to the kernel for performance optimization) to clear the contents of specific memory pages, even when MPK protects these pages from modification. Code can also employ the \textit{brk} and \textit{sbrk} system calls (originally designed for heap data management) to reclaim and reallocate sensitive data on the heap.\cite{connor2020pku} Therefore, security mechanisms should encompass the capability to intercept and filter privilege-level switches, such as system calls, in addition to functionalities like data isolation, boundary checks, and control flow restrictions. This is essential to prevent attacks exploiting privilege-level switches.

\section{Design of DASICS}

Given the challenges of designing secure processors, we aim to develop an efficient, minimally intrusive, and portable technique for enhancing processor security in applications sharing the same address space. This technique considers data flow integrity, control flow integrity, and system call integrity.

\textbf{Threat model:} We can categorize the internal software into two distinct components: developer-authored code segments and third-party libraries. The former can be managed and recompiled following new requirements by the software developer. The latter may consist of closed-source binary code or dynamically generated code for which source code analysis and developer modification are not feasible. Given that third-party libraries are susceptible to containing memory access vulnerabilities and potentially hazardous operations, such as privilege elevation via system calls, we assume that these third-party libraries are non-trusted. Attackers could potentially exploit these vulnerabilities to launch various attacks, including control flow hijacking, data manipulation, and the theft of critical data.

\textbf{Code-as-subject:}
To establish isolation within a shared address space, it is imperative first to address identifying authoritative subjects. On the one hand, conventional processes and threads within a single address space are unsuitable as authoritative subjects. On the other hand, selecting libraries, pointers, or instructions as authoritative subjects may result in overly broad or fine-grained granularity. We observe that code segments, constituting a contiguous sequence of addresses within the same program, can serve as a suitable unit for privilege segmentation. For instance, numerous code segments with consecutive addresses often exist within a library function, collectively contributing to implementing the same function. By ascertaining the extent of data accessed by each code segment at a specific point in time and the operations executed on that data, we can spatially segregate data manipulated by distinct code segments concurrently (e.g., constraining pointer access scope) and impose operational restrictions (e.g., confining read/write operations to specific addresses) to achieve security isolation at that particular moment. Moreover, it is essential to dynamically partition and restrict access as time progresses to guarantee secure isolation of the same code segment at different temporal instances (e.g., when the same function is invoked multiple times to process data at different locations). Furthermore, it is imperative to control the entry and exit points of these code segments (e.g., allowing functions to enter and exit exclusively through designated calls, without arbitrary jumps within the function or transitions to non-caller code at the conclusion), thereby regulating the control flow. Therefore, we choose code segments as the subject of the security decision, which we refer to as \textit{code-as-subject}. This aspect has also been mentioned in works such as CODOMs\cite{vilanova_codoms_2014}.

\textbf{Dynamic permission:} Based on the concept of code-as-subject, our design security mechanism sets the allowed access data boundary and jump target dynamically before invoking a code segment, relying on preset or presumed code data access range. The code segment can then freely access and jump within the specified region. An exception is triggered to block the access if an out-of-bounds access or illegal jump occurs. Permissions are automatically revoked when the code segment returns from execution.

The requirements of dynamic permission segmentation manifest in various ways:
\begin{itemize}
    \item Dynamic allocation and configuration of permissions are imperative due to limited security resources or metadata, which is particularly crucial for intricate programs that demand the segmentation and isolation of numerous code segments.
    \item Varying permission configurations are required for the same code segment under different circumstances. For instance, a function invoked by trusted code may receive authorization to access private data during its initial execution. In contrast, when invoked by non-trusted code, its access to private data must be deactivated. 
    \item Dynamic permission segmentation is essential for addressing situations where static segmentation is unfeasible. For instance, when the caller lacks knowledge about the precise behavior and semantics of a function pointer, they cannot anticipate the extent of data access.
\end{itemize}

\textbf{Built-in trusted base:} The security metadata-related issue mentioned above highlights the necessity of relying on a trusted base to manage security resources. Consequently, we need a trusted base to centralize privilege management and metadata upkeep. This trusted base must exhibit specific fundamental characteristics. Firstly, it should ensure that it is invulnerable or, at the very least, highly resistant to attacks. Secondly, the authority should transfer efficiently from the trusted base to non-trusted code to minimize performance overhead. The essential attributes of the trusted base are as follows:

\begin{itemize}
    \item The trusted base must be able to allocate and release security resources efficiently, facilitating security provisions for diverse code segments or various phases of the same code segment, even when confronted with limited security resources, thus enabling dynamic partitioning.
    \item The trusted base must establish the data regions accessible to non-trusted areas and specify the operations to be executed before the control flow transitions to non-trusted regions, thereby ensuring the data flow integrity.
    \item The trusted base must limit its entry points to prevent non-trusted areas from exploiting control flow hijacking to access trusted areas and manipulate security metadata.
    \item The trusted base should generate exceptions to address situations where non-trusted code segments perform actions that contravene privilege restrictions, providing the choice to terminate or rectify the situation.
    \item The trusted base must furnish a collection of system call-like functions that enable secure invocations of data and functionality, surpassing what non-trusted code can access independently.
\end{itemize}

In implementing trusted bases, we opted not to employ the approach of subdividing privilege levels and privileged states within the same address space. We discovered that the trusted base can serve as a distinct code fragment. The transition between trusted and non-trusted bases does not necessitate a complex privilege level switch; instead, it can be executed directly under predefined security rules. The trusted base is essentially an integral component embedded within the application.

We term the secure processor design scheme described above, which revolves around code-segment-based dynamic permission segmentation, as \textit{Dynamic in-Address-Space Isolation by Code Segments} (DASICS). Notably, the DASICS scheme comprises area segmentation, code permission segmentation configurations, control flow restrictions, permission validation, permission exception handling, primary function invocations, and system call filtering.

\subsection{Code Segmentation}
\begin{figure}
\centering
\includegraphics[width=85mm]{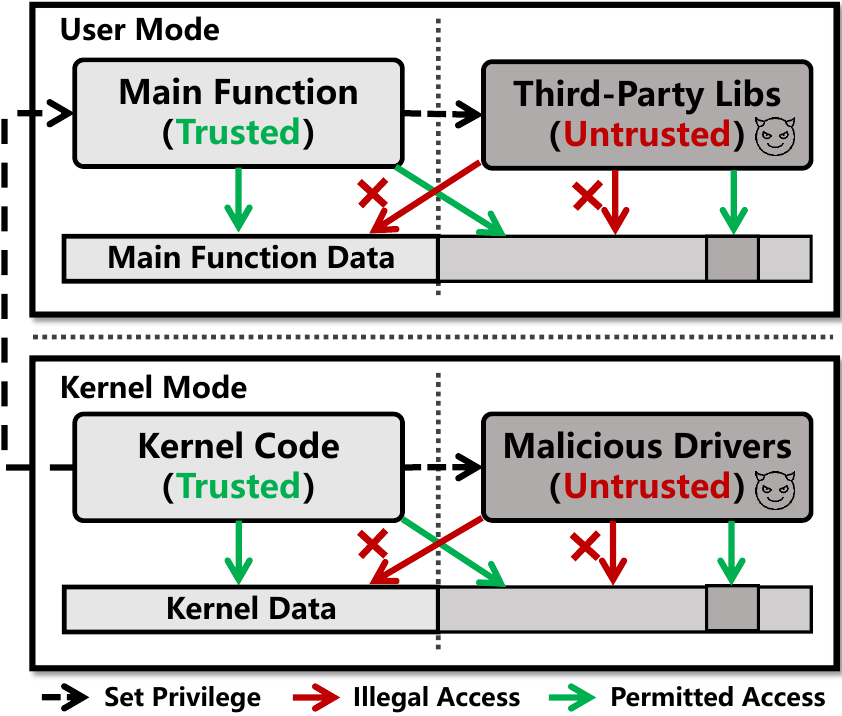}
\caption{DASICS Address Isolation (Horizontal/Vertical)}
\label{fig:isolation}
\end{figure}

DASICS divides privileges based on code segments, where a specific code segment refers to a set of executable codes located within a continuous address range (e.g., the code within a non-inline function). The boundary of a code segment can be determined by its start and end addresses. DASICS follows the principle that a code segment with higher privileges can restrict a code fragment with lower privileges. As illustrated in Figure \ref{fig:isolation}, these restrictions operate on two dimensions:
\begin{itemize}
    \item Vertical restriction: DASICS facilitates the configuration of privileges, allowing high-privilege codes to set and restrict low-privilege codes. For instance, in the RISC-V architecture, the trusted region within the M-mode can impose access restrictions on the trusted region within the S-mode, and correspondingly, the trusted region within the S-mode can constrain access to the trusted region within the U-mode.
    \item Horizontal Restriction: The DASICS methodology involves partitioning program code within the same privileged state into trusted and non-trusted zones. The trusted zone enforces access restrictions on the non-trusted zone. To illustrate, in the user mode, the trusted zone of the program typically comprises the main function, while the non-trusted zone includes third-party library functions. The main function configures privileges of the non-trusted zone before invoking the library functions. Meanwhile, in the kernel, the trusted zone encompasses the kernel scheduling codes, whereas the non-trusted third-party driver codes reside within the non-trusted zone. The kernel configures the necessary permissions before utilizing these non-trusted drivers.
\end{itemize}

Currently, the segmentation between trusted and non-trusted regions is established through manual address space marking. During the linking process, DASICS allocates trusted code segments to the statically defined address space to create the trusted zone. Codes residing in other address spaces are classified as non-trusted. Trusted zones possess unrestricted access to data at the same privilege level, whereas non-trusted zones are restricted from accessing data within trusted zones. Consequently, this delineation accomplishes segregating access areas for code segments with differing privilege levels.

\subsection{Code Privilege Segmentation}
DASICS categorizes code segment permissions into access and jump permissions. Access permissions are established based on data access boundaries. DASICS augments the hardware with multiple sets of boundary registers, each defining the extent of an accessible region and the allowable operations (read/write) within that region. Similarly, jump permissions employ jump range registers to specify the permissible target address ranges to which library functions can jump.

Before invoking codes within the non-trusted zone, the trusted zone takes the initiative to allocate boundary registers and configure permissions for the non-trusted code. These permissions encompass access and jump permissions, constraining data flow and control flow within the non-trusted zone. After the control flow returns from the non-trusted zone to the trusted zone, DASICS will reclaim the boundary register resources. In terms of control flow restrictions, in addition to defining the accessible area within the non-trusted zone through jump range registers, the entry points allowing jumps from the non-trusted zone to the trusted zone are also constrained. Specifically, the non-trusted code can only transition to trusted code through designated entry points such as function returns, main function calls, and same-privilege-level interrupts. This measure serves to prevent interception and exploitation of the trusted zone code.

\subsection{Privilege Check}\label{sec:priv-check}
The DASICS access privilege check is performed within the processor's fetch and access units. When non-trusted code performs access operations (including fetch/read/write), DASICS will automatically examine the access address against the boundary register. Suppose an access operation beyond the permitted privilege level is detected, such as an out-of-bounds address or an unauthorized operation. In that case, DASICS triggers an exception at the same privilege level for further handling (e.g., exiting with errors).

Control flow restriction checks are conducted during the execution of jump instructions. When the control flow transitions from a trusted zone to a non-trusted zone via a function call, DASICS will record the function's return address within the trusted zone. Upon returning from the non-trusted zone, DASICS will compare the actual return address with the recorded return address. If they do not match, DASICS will raise an exception to indicate that the non-trusted code may tamper with the return address.

Leveraging the privilege checks mentioned above, DASICS can mitigate control-flow hijacking attacks, including buffer overflows and malicious modifications of return addresses.

\subsection{Same-Privilege-Level Exception Handling}

DASICS supports the trusted zone to register exception handling functions for access exceptions, control-flow exceptions, and system call filtering. In line with the privilege checking in section \ref{sec:priv-check}, when an exception occurs, it triggers an interrupt at the same privilege level (e.g., triggering a user-level interrupt in the user mode). This interrupt leads the program to transfer to the exception handling function, retrieving the specific reason for the exception from the hardware privilege register. Subsequently, the exception handling function undertakes distinct actions based on the specific cause of the exception, such as terminating with an error or proceeding with the original instruction execution after dynamically correcting the privilege. Moreover, the exception handling function supports dynamic registration, accommodating varying handling procedures for different exceptions and scenarios.

\subsection{Trusted Call}

In a manner analogous to the system calls of an operating system, DASICS restricts non-trusted zone data access while also providing trusted calls for non-trusted zones to establish boundaries for DASICS access. For instance, library functions within the non-trusted zone utilize trusted calls to revoke their private data access privileges, thus preventing unauthorized access from other non-trusted zone functions. It is important to note, however, that a non-trusted zone can only employ trusted calls within the confines delineated by the trusted domain. Non-trusted zone functions are prohibited from elevating their own privileges through trusted calls, and when initiating a trusted call, they are subjected to parameter validation to ensure compliance with the defined boundaries.

To support trusted calls, the trusted domain must first register the entry points for trusted calls to expand the legitimate entry points within the trusted domain. Non-trusted zone code can configure trusted call parameters in a manner akin to system calls and subsequently execute a direct jump to these entry points for trusted calls. Notably, this approach is more expedient than system calls, as it obviates the time-consuming privilege level transitions.

\subsection{Syscall Filtering}
To counteract attacks that exploit security-sensitive system calls, DASICS incorporates a system call filtering mechanism. Whenever the non-trusted zone executes a system call instruction (e.g., the \textit{ecall} instruction in the RISC-V architecture), the monitoring logic deployed by the hardware will trigger a user-level interrupt and redirect the system call to an exception handler within the trusted zone. Subsequently, the exception handler transfers control to the appropriate system call exception handler, depending on the nature of the exception (e.g., user-state system call exception). This function can conduct security checks for this system call, such as verifying whether the code segment is authorized to execute the specific system call and assessing whether the system call's parameters fall within the permission range of the code segment. If the system call passes the above checks,  the exception handler function will finally emulate it. Implementing system call checking within the application further allows for precise and granular management of various file and device resources.

\subsection{Advantages of the DASICS solution}
When compared to prior security protection mechanisms addressing memory access vulnerabilities, we posit that the DASICS scheme offers the following advantages:

\begin{itemize}
    \item \textbf{Comprehensive protection}: DASICS ensures data flow integrity (DFI) by implementing memory isolation for subjects with varying privileges and control flow integrity (CFI) through pairwise validation of function calls and returns. Besides, DASICS can safeguard application data and significantly reduce the likelihood of unauthorized modifications to the security metadata.
    \item \textbf {Efficiency}: In DASICS, the transition between trusted and non-trusted regions is executed through a direct jump, eliminating the need for context switches or interrupts as observed in mechanisms like Donky. Furthermore, this approach minimizes the necessity for frequent interactions with the kernel or secure monitors, culminating in a comprehensive and efficient security protection mechanism.
    \item \textbf{Fine-grained protection}: DASICS offers a finer granularity in protection when contrasted with page-grained protection mechanisms like MPK. It achieves this by dynamically partitioning the access region using boundary registers to confine pointer access, resulting in more precise permissions.
    \item \textbf{Dynamic protection}: DASICS accomplishes dynamic segmentation of memory space by dynamically defining trusted zones and allocating/reclaiming security resources (such as access boundary registers and jump range registers) to achieve adaptable protection tailored to specific requirements. This capability also enables the enforcement of timely security rules in various scenarios.
    \item \textbf {Multi-level privilege protection}: DASICS supports security protection on multi-level privilege modes, encompassing kernel and user modes. We can apply it not only for safeguarding code in the user mode (refer to Section \ref{subsec:case1}) but also in the kernel mode (refer to Section \ref{subsec:case2}).
    \item \textbf {Fine portability}: Thanks to the DASICS protection mechanism, users can only designate a non-trusted zone for third-party libraries without recompiling them. When the source code can be modified, we can achieve additional security enhancements by introducing a straightforward API at the function entry and exit points.
\end{itemize}

\section{Implementation}
We have successfully realized a prototype demonstrating the cooperative functionality of DASICS between hardware and software, following the design principles outlined above. In the hardware domain, we have implemented DASICS on two domestically developed RISC-V architecture processors: the NutShell\cite{NutShell} in-order processor and the XiangShan\cite{XiangShan} out-of-order superscalar processor. On the software front, we have created a DASICS system-level simulator based on QEMU/NEMU and have modified DASICS system software using the Berkeley BootLoader (BBL) and Linux 4.18.2 as the foundation. In the future, our efforts will continue to encompass porting tasks, aligning with OpenSBI, U-Boot, and the Linux kernel version 5.10.

\subsection{Hardware Prototype}
\begin{figure}
\centering
\includegraphics[width=88mm]{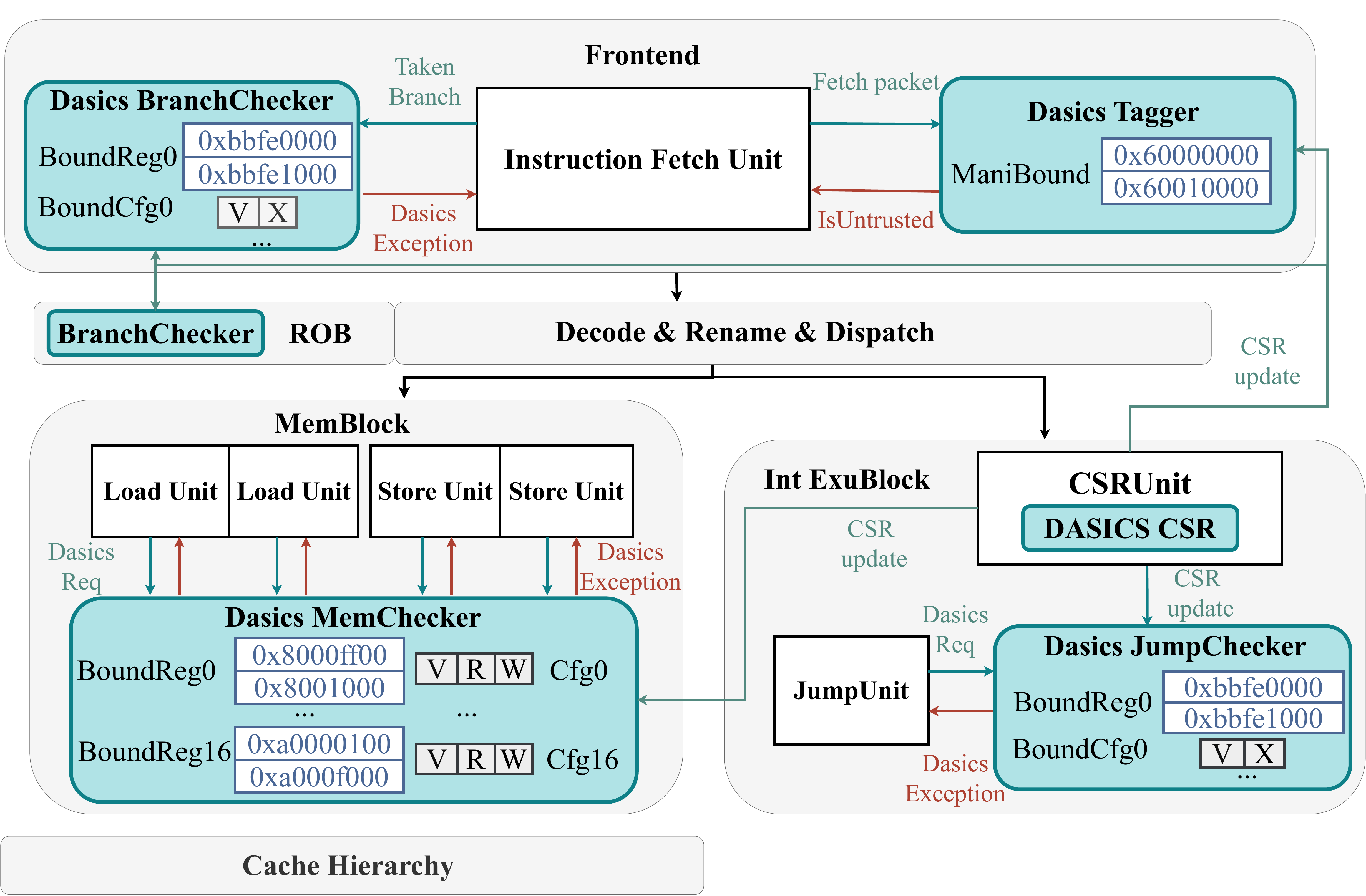}
\caption{DASICS modules on XiangShan}
\label{fig:xs-dasics}
\end{figure}

We have implemented hardware prototypes of DASICS in both the NutShell in-order open-source processor and the XiangShan out-of-order superscalar processor. Taking the XiangShan processor as an example, as illustrated in Figure \ref{fig:xs-dasics}, we have utilized RISC-V CSR registers to enable software-defined trusted code address ranges. The software can configure these boundary registers through CSR read and write instructions. Within the CSR unit, we have also incorporated functionalities such as DASICS protection toggles, trusted call entry addresses, and function call return address storage. Additionally, we have integrated a DASICS-Tagger module in the processor's front end to determine whether the current instruction stream belongs to the trusted code based on the defined trusted code address ranges, and we have incorporated DASICS checking modules in the memory access unit and branch unit modules within the XiangShan processor core's execution backend.

In the memory access unit, we have implemented memory boundary checks based on CSR (Control and Status Register) registers. Trusted code limits the data regions that non-trusted code can access by configuring these registers. Non-trusted code's memory access instructions (load/store) perform address translation using the TLB (Translation Lookaside Buffer) while conducting virtual DASICS memory access checks simultaneously. If a memory access attempt goes beyond the defined boundaries or violates access permissions, it triggers the corresponding DASICS user-mode exception. The memory access unit can accommodate 16 boundary registers, each comprising upper and lower virtual address bounds for the specified range, an effective flag, and permitted operations (read/write).

Within the jump unit of XiangShan, we have also implemented memory boundary checks based on CSR registers. Trusted code can ensure control flow integrity by configuring the jump targets for non-trusted code. Specifically, non-trusted code is only allowed to perform the following types of jumps:

\begin{itemize}
     \item The jump target is the return address of a trusted code function call.
     \item The jump target is a trusted entry point specified by the trusted code.
     \item The jump target falls within a specified range of code segments within the trusted code, referred to as the \textit{active zone}.
\end{itemize}

Simultaneously, we have imposed restrictions on conditional branch instructions, limiting the target addresses of non-trusted code's branch instructions to fall within the active zone defined by the trusted code. This measure is taken to prevent potential attackers from using non-trusted code's conditional branch instructions to bypass the checks in the branch unit and gain direct control over the trusted code for unauthorized modification of secure metadata. The branch unit and the front end have been augmented with checks for target addresses in jump instructions (jal/jalr) and conditional branch instructions (branch). These checks include four sets of boundary registers, each analogous to those used in memory access, encompassing upper and lower address bounds for the specified range and an effective flag.

\begin{figure*}
\centering
\includegraphics[width=150mm]{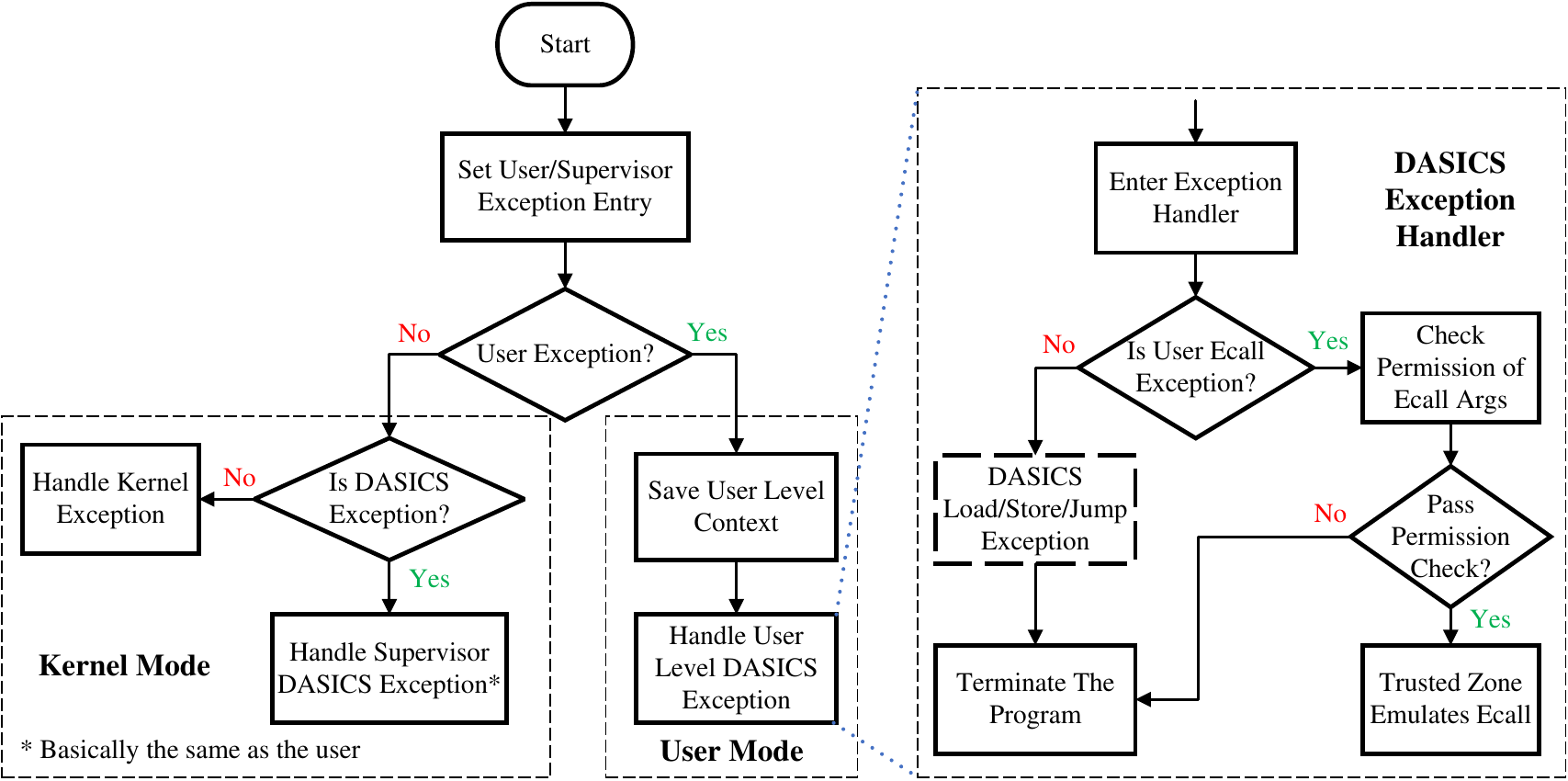}
\caption{DASICS Exception Procedure}
\label{fig:exception}
\end{figure*}

To handle corresponding DASICS violations, we have introduced new DASICS exceptions. These exceptions encompass memory access violations, jump violations, and DASICS system call interception exceptions. We have implemented triggering mechanisms for these exceptions in both S-mode and U-mode and have also implemented RISC-V architecture N-extensions to support user-mode interrupts. For example, the \texttt{ecall} instruction in non-trusted code triggers the DASICS exception in U-mode, redirecting control to the DASICS exception handler. Within this handler, non-trusted code system call parameters are examined and intercepted, including but not limited to checking system call parameters against non-trusted code's authorized memory access and jump ranges to prevent out-of-bounds system calls. The specific exception-handling process is illustrated in Figure \ref{fig:exception}.

To achieve the vertical and horizontal constraints described in the DASICS design, we have established that the boundaries of the trusted zones in U-mode can only be set by the trusted zones in S-mode, and the boundaries of the trusted zones in S-mode can only be set by the trusted zones in M-mode. Within the same privilege level, CSR registers related to DASICS can only be configured by the trusted code. Any attempt by non-trusted code to access DASICS triggers an illegal instruction exception.

In the six-stage pipeline structure of the NutShell in-order core, the modules we introduced primarily include the instruction decode unit (IDU), CSR execution module, and the load-store unit (LSU), among others. The total number of lines of Chisel code modified did not exceed 500. We conducted testing under hardware simulation using Verilator and on FPGA development boards (NutShell was tested on the Xilinx PYNQ-Z2 development board, and XiangShan was tested on the Xilinx VCU128 development board).

\subsection{Software Prototype and Support}
We have implemented the DASICS software simulator based on QEMU version 8.1.0. Additionally, we have implemented the DASICS mechanism using the NEMU simulator to support XiangShan's difftest verification framework, ensuring the correct RTL-level implementation of DASICS.

For system software, we have also made preliminary implementations of the DASICS mechanism. In BBL, we have added the initialization of the DASICS mechanism, including setting up proxies for DASICS exceptions and configuring the S-mode trusted zone to encompass the entire memory space. After BBL completes initialization, it jumps to the S-mode trusted zone, where the S-mode trusted zone, through newly added SBI calls, modifies the trusted zone boundaries.In Linux, we have added DASICS exception handling functions and incorporated the DASICS mechanism into the ELF loading of user programs. Specifically, user software explicitly specifies the trusted and non-trusted zones, which is encoded in the ELF file. During loading, the operating system automatically configures the program's initial trusted zone and enables the DASICS switch based on the DASICS partition information in the ELF file, using CSR instructions. During runtime, trusted zone functions, before calling non-trusted zone functions, first use CSR instructions to set the accessible range and jump range, thereby restricting the permissions of non-trusted zone functions. When non-trusted code violates these restrictions, it triggers a user-mode interrupt, entering the DASICS user-mode exception handler for the corresponding operation. Furthermore, we provide the DASICS software library, which includes trusted call code, exception handling code, and system call verification code.

\section{Protection Case Study}
\subsection{Case Study 1: Protection of Third-Party Untrusted Library}\label{subsec:case1}
\begin{figure}[ht]
\centering
\includegraphics[width=88mm]{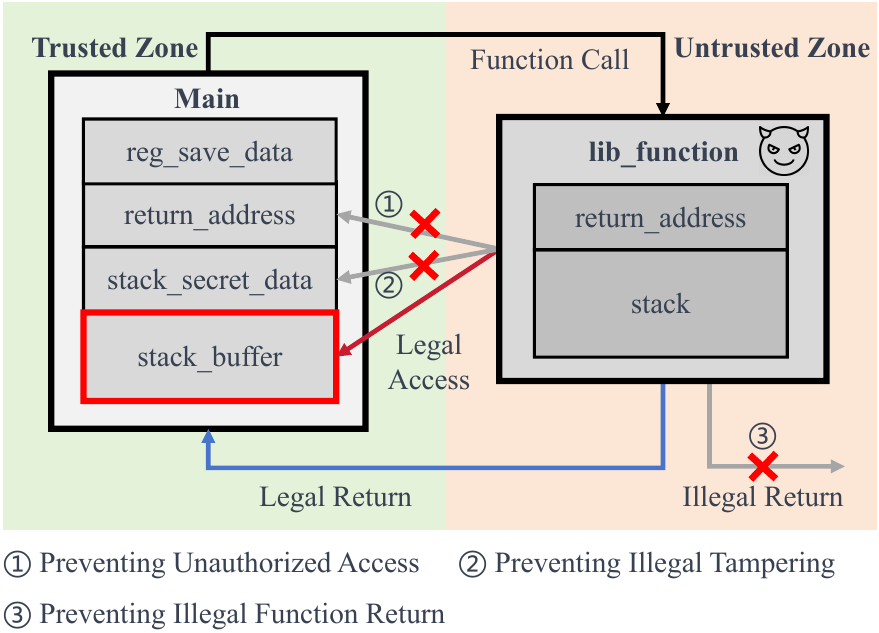}
\caption{Protection of Third-Party Untrusted Library Functions}
\label{fig:case-study-1}
\end{figure}

We illustrate the protection offered by DASICS for both data flow and control flow through a practical example within the user-mode address space. In this example, the main function of a user program, denoted as \textit{main}, makes use of a third-party untrusted library function called \textit{lib\_function} to perform certain computations. As depicted in Figure \ref{fig:case-study-1}, the main function divides a segment of its stack space into three distinct regions: an array used for computation (\textit{stack\_buffer}), a private data region specific to the main function (\textit{stack\_secret\_data}), and a register save area (\textit{reg\_save\_data}).

The main function passes the address of \textit{stack\_buffer} to the untrusted library function \textit{lib\_function} as an argument for array computation. However, the \textit{lib\_function} contains malicious code that illicitly employs \textit{memcpy} to overstep its bounds and steal data from both \textit{stack\_secret\_data} and \textit{reg\_save\_data}. Even more critically, the malicious code manipulates the return address within the \textit{reg\_save\_data} region, diverting the normal program flow. Consequently, when the main function returns, it does so abnormally and transfers control to a target address set by the malicious code, enabling further exploitation.

In response to this scenario, we employ DASICS data flow protection. Initially, the main function is designated as trusted code, while the \textit{lib\_function} function is classified as untrusted code. Before the main function calls \textit{lib\_function}, we employ DASICS to restrict \textit{lib\_function} to only access its own stack space and the array used for computation (\textit{stack\_buffer}). This way, any attempt by malicious code within \textit{lib\_function} to access the main function's private data region (\textit{stack\_secret\_data}) through out-of-bounds access triggers a user-mode DASICS memory read exception, resulting in termination. Similarly, when malicious code attempts to modify the return address in the register save area (\textit{reg\_save\_data}), it triggers a user-mode DASICS memory write exception.

However, relying solely on the aforementioned data flow protection is insufficient. This is because the 'lib\_function' function could potentially modify its own return address and then return to specific code within the main function, enabling data theft or tampering. In more extreme cases, post-return control flow manipulation, such as ROP attacks or DOP attacks\cite{prandini2012return,hu2016data}, could be employed to hijack DASICS configuration functions, thereby elevating the attacker's data access privileges and circumventing DASICS data flow protection.

In response to the second scenarios, we employ DASICS control flow protection. When the main function calls \textit{lib\_function}, it utilizes a special jump instruction, \textit{dasicscall.jr}, for the transition. In addition to the conventional control flow transfer semantics of regular jump instructions, this instruction also stores the return address (the address of the next instruction in sequence) in the \textit{dasics\_ret\_pc} CSR (Control and Status Register).

Upon the return of the library function (through the \textit{ret} instruction), a check is performed to verify whether the return address matches the value stored in \textit{dasics\_ret\_pc}. If there has been any tampering with the return address within \textit{lib\_function}, it triggers a user-mode DASCIS jump exception. This effectively prevents the library function from bypassing data flow checks through control flow manipulation.

\subsection{Case Study 2: System Call Interception in User Space}\label{subsec:case2}
In our experiment involving system call interception for a frequently used command-line tool \textit{ls}, on Unix and Unix-like operating systems, we have augmented the main function's source code of \textit{ls} to incorporate the relevant system call interception logic. Specifically, on the Linux operating system platform, we intercept and scrutinize the following three major system calls:

\begin{itemize}
     \item \textbf{\textit{openat}}: This system call is used to open a file in the directory represented by the specified file descriptor. It is commonly used for file operations involving relative paths.
     \item \textbf{\textit{getdents}}: This system call is employed to retrieve directory entry information from a directory. It is typically used for enumerating files and subdirectories within a directory to list its contents.
     \item \textbf{\textit{fstatat}}: This system call is utilized to obtain status information (stat information) for a specified file in a given directory.
\end{itemize}

After the appearance of system call instructions within \textit{ls}, the DASICS system call user-space exception is triggered, redirecting control flow to a processing function, this processing function then diverts control flow based on the system call number. The aforementioned three system calls, as indicated, print their respective parameters. Through this approach, we can monitor non-trusted region system calls and impose restrictions on critical parameters.

Our work achieves the logic for monitoring system calls throughout the entire program with minimal modifications solely to the program's main function. Furthermore, we can implement additional functionalities such as parameter validation and system call filtering. Existing literature on system call interception indicates that software-based system calls interception, such as \texttt{strace}, based on \texttt{ptrace} and \texttt{seccomp}, incurs significant overhead. Therefore, we contend that our user-space exception-based system call interception is more efficient than software-based implementations.

\section{Future Application of DASICS}

\subsection{Use Case 1: Rust + Untrusted C/C++ library}

As mentioned, programming with safety-oriented languages like Rust can address most memory vulnerabilities. However, developing in such secure programming languages is intricate, and porting existing codebases to these languages entails substantial effort. With the integration of DASICS, developers can construct trusted enclave code using secure languages while imposing restrictions on other untrusted library code through the DASICS mechanism. This approach ensures not only the absence of memory vulnerabilities in the trusted enclave code but also significantly reduces the workload of rewriting existing code. Permissions for untrusted code in the non-enclave region are constrained within the trusted enclave through the DASICS mechanism, thereby mitigating potential security risks in the non-enclave region. DASICS achieves security privileges by manipulating hardware registers, making it language-agnostic and adaptable to various programming languages, offering flexible security configurations.

\subsection{Use Case 2: Kernel + Untrusted Third-Party Drivers}
Many existing security mechanisms rely on a secure kernel, which enforces security checks or restrictions on user-space programs by configuring certain security privilege data within the kernel. However, the kernel itself may run a substantial amount of third-party code, including file systems, device drivers, network protocol stacks, and other kernel modules. These modules, in the eyes of the operating system developers, are not inherently trustworthy, leading to security vulnerabilities. Both third-party modules and kernel code execute at a privileged level, and once the security of the operating system kernel itself is compromised, many existing security defenses become ineffective.

Through the DASICS mechanism, we can introduce a separation between trusted and untrusted regions within the operating system as well. System developers can place third-party module code in the untrusted region of the kernel space and restrict their execution and access privileges, thereby mitigating security risks introduced by third-party kernel modules. The DASICS mechanism leverages hardware to implement security protection, enabling the isolation and protection of kernel addresses without undermining the existing separation between kernel mode and user mode. Additionally, it introduces a new level of isolation within the kernel mode.

\subsection{Use Case 3: Unikernel (Library OS)}
In scenarios such as cloud computing, applications often do not require the complex functionalities of a full-fledged operating system. However, a complex operating system can introduce overhead and reduce overall system efficiency. As a result, the concepts of LibraryOS and Unikernels have been proposed. These concepts transform the necessary operating system functionalities for applications into user-space libraries, thereby reducing the overhead imposed by complex operating systems and enhancing platform adaptability for applications. However, such solutions, while achieving performance gains, often compromise the traditional security provided by the isolation between kernel mode and user mode.

To attain sufficient security, additional measures are required to implement lightweight isolation mechanisms. By incorporating DASICS functionality into Library OS, it becomes possible to restrict memory access permissions for user programs and enforce these restrictions at the function granularity. This isolation effectively separates the code for application and kernel functionalities within Library OS, mitigating certain security risks. Compared to protection mechanisms relying on multiple privilege mode switches, DASICS incurs lower performance overhead, thereby preserving the original performance advantages of Library OS.

\section{Discussion}

In recent years, with the increasing emphasis on efficient security protection through technologies such as WebAssembly, Unikernels, and third-party libraries, there has been a growing recognition of systematic research into memory safety issues within the same address space. Many methods and techniques for isolating address spaces share certain similarities with DASICS, but they also exhibit differences in certain approaches and philosophies. In this chapter, we will elucidate the distinctions between DASICS and these methods.

The precursor research for DASICS stems from IMPULP\cite{zhao2020impulp}, a technology that employs code segment partitioning for runtime protection of user-space address spaces. DASICS builds upon the ideas and implementation of IMPULP. Firstly, DASICS extends the application scenarios beyond those of IMPULP. Originally, IMPULP focused solely on protecting user-space address spaces. As mentioned earlier, DASICS extends its protection to user-space and kernel-space, encompassing different system privilege levels. Secondly, DASICS enhances the capability for control flow checks compared to IMPULP. IMPULP primarily examined function return addresses, while DASICS further leverages jump boundary registers to confine the code range where control flow transitions can occur in non-trusted regions. This further fortifies the program's defense against attacks such as Return-Oriented Programming (ROP). Moreover, we employ the RISC-V N extension\cite{RISCV-N} to support user-space interrupts and implement user-space handling of DASICS exceptions based on this extension, which is more efficient compared to the exception handling within the kernel in IMPULP. IMPULP's hardware prototype was based on the RocketChip in-order core, while DASICS extends this work to the XiangShan out-of-order core implementation. Additionally, on the software front, DASICS introduces extensions for dynamic linking, trusted calls, and simple nested calls to non-trusted region functions.

The concept of permission delineation based on code was introduced as an innovative aspect in CODOMs\cite{vilanova_codoms_2014}. CODOMs initially divide data regions within the same address space into distinct permission entities, such as different functions, at the granularity of pages. Simultaneously, CODOMs keep records of what actions a particular permission entity can perform on other permission entities, including function calls and data access. The criterion for partitioning permission entities is based on code addresses, and the practical implementation involves adding labels to all pages belonging to the same permission entity. Typically, only code with the same label can access data with the corresponding label. Access across permission entities is restricted, and CODOMs use a table to record permitted inter-permission-entity operations in the actual design. Furthermore, CODOMs address the granularity issue in data protection, recognizing that page-level protection alone is insufficient. This limitation can hinder small permission entities (e.g., functions) from efficiently performing fine-grained data permission granting operations. Hence, CODOMs propose a method for rapid permission propagation based on Capability registers, with the design and implementation principles of these registers resembling the boundary registers in DASICS. CODOMs conducted comprehensive deliberations on the entire system of code-based protection. However, there are certain limitations. On one hand, crucial permission-related data structures in CODOMs, such as page labels and inter-permission operations tables, are maintained by the operating system. This oversight exposes CODOMs to threats originating from the operating system and entails overhead in terms of safeguarding metadata updates due to privilege mode transitions. On the other hand, since CODOMs were published quite early, they lack mechanisms for intercepting and inspecting system calls, leaving potential bypasses that could compromise security protection.

The compartmentalization between trusted and untrusted regions was not originally proposed by DASICS; Donky\cite{schrammel2020donky} is a work that enhances defense using Memory Protection Keys (MPK). In the design of Donky, each permission entity is represented by a cluster of MPK keys to delineate its scope of permissions. The creation of each permission entity, assignment of keys, and inter-permission-entity operations all require handling within a component known as the Domain Monitor, which is a runtime program segment operating in user mode. It is evident that this concept aligns closely with DASICS' notion of trusted regions, where all permission management for untrusted code is centralized within the trusted region for allocation and validation. Furthermore, Donky, much like DASICS, utilizes the RISC-V N extension\cite{RISCV-N}, specifically user-space interrupts, for intercepting and inspecting system calls. However, in contrast to DASICS, Donky's transitions between permission entities are more intricate. Taking function calls as an example, the caller needs to explicitly trigger a user-space interrupt using a specific instruction, leading to entry into the Domain Monitor. Within this context, the process involves preserving registers, validating function calls and their parameters, and updating keys for the callee. Similarly, the return from a function call follows a similar procedure. This intricacy results in a substantial overhead for each function call in Donky, as all checks depend on the Domain Monitor. Moreover, Donky inherits the issue of excessive protection granularity inherent in MPK-based approaches, similar to DASICS. However, Donky restricts the use of update instructions for the PKRU register solely within the Domain Monitor, which enhances its security compared to conventional MPK methods.

SecureCell\cite{bhattacharyya_securecells_2023} presents a more systematic and comprehensive examination of address space isolation methods, delving into aspects such as security, performance, and scalability. They outline 11 specific design objectives and propose a security architecture based on permission tables in accordance with these objectives. In SecureCell, each permission entity is assigned a unique ID, and contiguous blocks of data or code are divided into cells. SecureCell modifies the page table to enable the virtual memory system to support the delineation of permissions (read/write/execute) for each cell by each permission entity. SecureCell also supports sharing, allowing each cell to have different permissions for different permission entities. Importantly, SecureCell facilitates explicit permission transfers between different permission entities through code, decoupling the transfer into granting and receiving phases. A cell's permission transfer can only be completed if the granter agrees to grant and the receiver is willing to accept, enhancing security compared to one-way transfers that only consider granting. This approach mitigates security concerns associated with forced grants, such as code injection attacks. Although SecureCell transitions from a page-based TLB structure to a range-based TLB structure, improving TLB storage efficiency, it remains confined to page-level protection. Furthermore, SecureCell implements a hardware prototype on the RocketChip in-order core, without consideration for issues specific to out-of-order processors.

HFI\cite{narayan2023going} is an optimization effort targeted at WebAssembly application scenarios. They highlight the shortcomings of software-based boundary checks and attempt to improve them through a combination of software and hardware. The architecture design of HFI closely resembles that of DASICS, incorporating instructions and related registers for entering and exiting the HFI protection mode within the processor. In the HFI protection mode, data access and code execution are both constrained by a set of boundary registers, referred to as \textit{Regions}. The boundary checking process is synchronous with TLB virtual-to-physical address translation, ensuring it does not introduce new pipeline stages. These boundary registers consist of a base address, data region size, and data permissions (read/write/execute). When an illegal access occurs, an exception is triggered, redirecting control to the user-mode runtime for handling. This mechanism shares similarities with DASICS' boundary register approach. HFI also captures and inspects system calls. Their method involves modifying the microcode of x86 system call instructions, altering the instruction semantics to direct execution to a specific system call checking function within the runtime. However, the application scenario for HFI dictates that it does not delve further into addressing the issue of supporting different permission entities under the HFI mode. In contrast, DASICS aims to resolve the problem of nested function calls in the untrusted region. It's worth noting that HFI was simulated on the gem5 simulator using an out-of-order x86 architecture and does not possess a physical hardware prototype.

\bibliographystyle{IEEEtranS}
\bibliography{ref}
\newpage

\section{Appendix: DASICS Open Source Code Repository}

\subsection{Hardware}
\begin{itemize}
     \item \textbf{\textit{NutShell}}: https://github.com/DASICS-ICT/NutShell-DASICS
     \item \textbf{\textit{XiangShan}}: https://github.com/DASICS-ICT/xiangshan-dasic (will release soon)
\end{itemize}

\subsection{Software}
\begin{itemize}
     \item \textbf{\textit{Linux}}: https://github.com/DASICS-ICT/riscv-linux
     \item \textbf{\textit{bbl}}: https://github.com/DASICS-ICT/riscv-pk
     \item \textbf{\textit{OpenSBI}}: https://github.com/DASICS-ICT/OpenSBI
     \item \textbf{\textit{QEMU}}: https://github.com/DASICS-ICT/QEMU-DASICS
     \item \textbf{\textit{NEMU}}: https://github.com/DASICS-ICT/NEMU
     \item \textbf{\textit{Case-1}}: https://github.com/DASICS-ICT/DASICS-case-study
     \item \textbf{\textit{Case-2}}: https://github.com/DASICS-ICT/DASICS-ls-protect
     \item \textbf{\textit{dynamic-link library}}: https://github.com/DASICS-ICT/DASICS-dynamic-lib
\end{itemize}

\end{document}